\documentclass[final,5p,times,twocolumn]{article}

\newcommand{\epem}              {\ensuremath{\mathrm{e^+e^-}}}

\usepackage{graphicx}

\usepackage{rotate}

\usepackage{amssymb}

\begin{document}




\title{DEPFET Pixels as a Vertex Detector for the Belle II Experiment}


\author{Jochen Schieck for the DEPFET Collaboration \\
Ludwig-Maximilians-Universit\"at M\"unchen, Am Coulombwall 1, D-85748 Garching, Germany \\ and
Excellence Cluster Universe, Boltzmannstr. 2, D-85748 Garching, Germany}

\maketitle

\begin{abstract}
Currently the heavy flavour factory KEKB located at the KEK accelerator centre in Tsukuba, Japan, is being
upgraded to the Super KEKB factory, aiming for a substantially higher luminosity of $8 \times 10^{35}$ cm$^{-2}$s$^{-1}$.
This by a factor of 40 increased luminosity requires significant modifications of the Belle detector. The vertex
detector needs to be redesigned and, besides a four-layer silicon strip detector, a pixel detector based
on the DEPFET technology will be installed closest to the interaction point. The requirements for this
pixel detector, the DEPFET technology and the current status of the pixel detector will be presented in this paper.
\end{abstract}







\section{Introduction}
The Standard Model of particle physics (SM) is a very successful theory describing consistently
all particle physics measurements being observed up to now. However, it is believed that this theory 
is only an effective theory since some important questions still remain open e.g. the SM does
not include gravity and cannot explain the observed baryon asymmetry in the universe. 
The quest for physics beyond the SM is therefore the major aim of modern particle
physics experiments. This search can be performed with two orthogonal approaches: production
of new particles with high energy colliders (like the LHC at CERN) or through precision
experiments like Belle II at the Super KEKB collider. With precision experiments 
new particles and new forces can appear in loop corrections and possible
deviations from SM predictions provide a clear hint for physics effects beyond the SM. If the coupling
to these new particles is large, the sensitivity to physics beyond the SM can be significantly
increased compared to direct detection experiments like LHC.  

\section{The Super KEKB Collider and Belle II Detector}
The Super KEKB collider is an asymmetric \epem-collider (e$^{-}$: 7 GeV, e$^{+}$: 4 GeV) 
operating at the $\Upsilon$(4s) resonance. The operation is expected to start in 2015 and 
will  provide 50 fb$^{-1}$ during 10 years operation time. Compared to the KEKB collider
the boost $\gamma\beta$ is decreased leading to a deteriorate resolution for time dependent
measurements. Several parts of the Belle detector have to be adapted to the new
Super KEKB running conditions. A detailed description of the planned upgrade including
a detailed summary of the pixel detector can be found
in~\cite{Abe:2010sj}. The vertex detector of the upgraded Belle detector ("Belle II") will undergo a major revision and
next to the silicon strip detector (SVD) a pixel detector (PXD) will be installed close to the interaction point. 
The increased luminosity will be accompanied by worse background conditions and 
the new pixel detector has to be able to cope with these conditions. In order to digest
the high background, the detector has to provide a fast readout cycle,  a high signal-to-noise ratio
and the detector has to be tolerant against irradiation.
To fulfil  the physics requirements the detector has to have an excellent hit resolution and 
built with a very low material budget. A pixel detector based on the DEPFET (Depleted p-channel
field effect transistor) technology~\cite{Kemmer:1986vh} fulfils all these requirements.

\section{The DEPFET Pixel Detector}

\subsection{The DEPFET Principle}
\label{depfetprinciple}
\begin{figure}[hbt] 
\centering 
\includegraphics[width=0.65 \columnwidth,keepaspectratio]{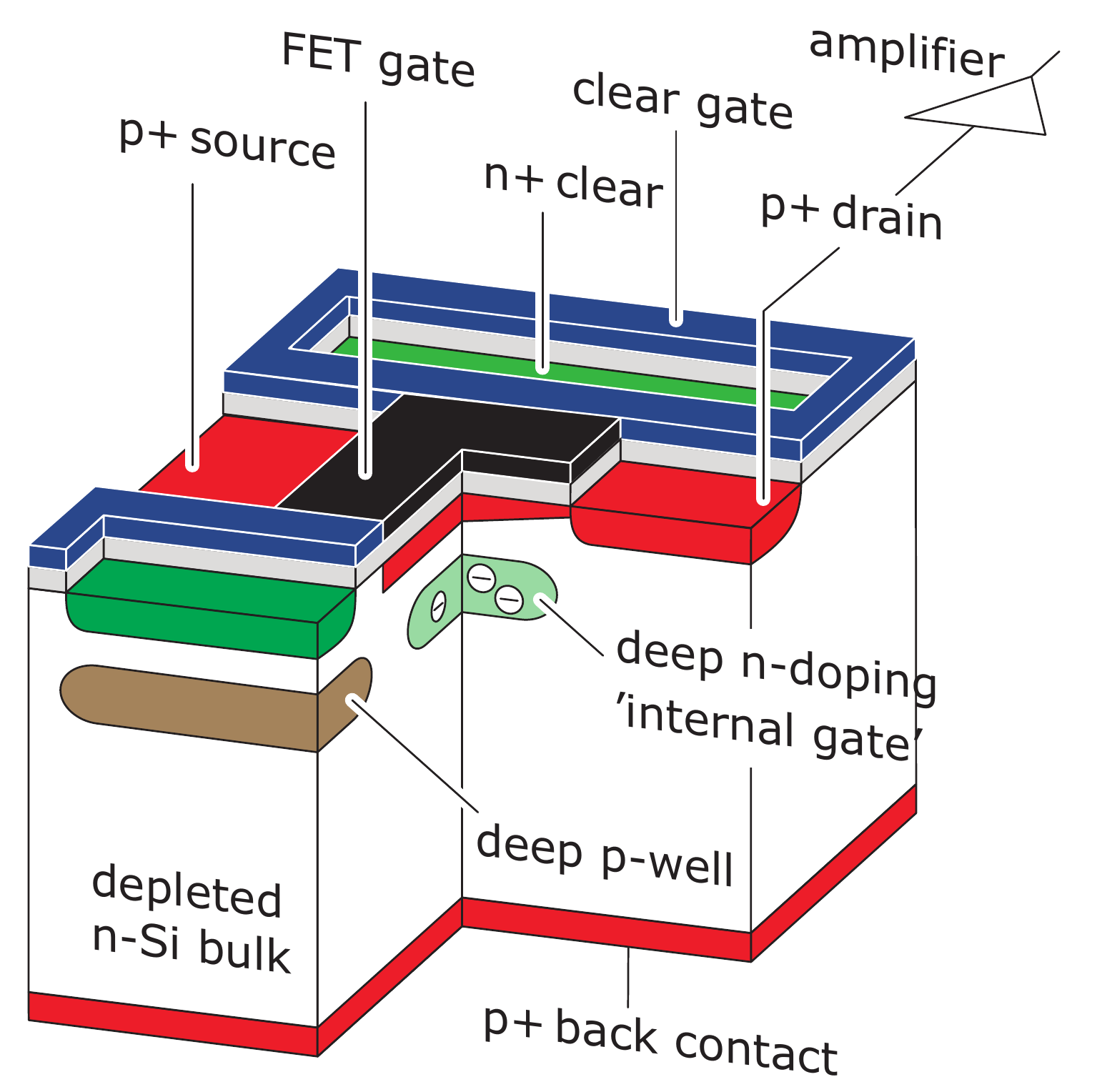}
\caption{Sketch of a single DEPFET cell taken from~\cite{Abe:2010sj}. Besides
the FET gate the transistor current is modulated by electrons collected in the internal
gate. The electrons can be removed via the clear contact.}
\label{fig:DEPFETCell}
\end{figure}

A single DEPFET pixel cell consists of a field effect transistor (FET) which is operated 
on sideways
depleted n-type bulk material. Electron-hole pairs produced by traversing particles 
are separated by the electric field and the electrons are collected in
the internal gate, a deep n-implant below the gate of the FET. The current 
in the field effect transistor between the source and the drain 
will then be modulated by the gate of the transistor and in addition
by the field generated from the electrons collected in the internal gate. After the 
measurement of the current the collected electrons are removed by the
 clear contact based on a 
punch-through mechanism to the internal gate. The layout of a 
single cell is shown in Fig.~\ref{fig:DEPFETCell} \par
A key parameter for the DEPFET performance  is the internal amplification parameter
$g_q$, which describes the change in transistor current $\Delta I_{D}$ divided by
the charge collected in the internal gate $\Delta q_{in}$: $g_{q}=\Delta I_{D}/\Delta q_{in}$. 
The expected number for the DEPFET sensors used in Belle II is $g_{q} \approx 400$ pA/e$^{-}$,
leading to a change in the FET current of 2.4 $\mu$A, assuming 6000 electron hole pairs
being produced in a 75 $\mu$m thick silicon layer.
The low power consumption is a major advantage of the DEPFET design. The detector
is always sensitive to charged tracks passing the detector, while the gate is only switched on
during the readout process. In the current design the active sensor area consumes only 
0.5 W per half ladder. 

\subsection{The Pixel Detector Layout}
\begin{figure}[hbt] 
\centering 
\includegraphics[width=0.72 \columnwidth,keepaspectratio]{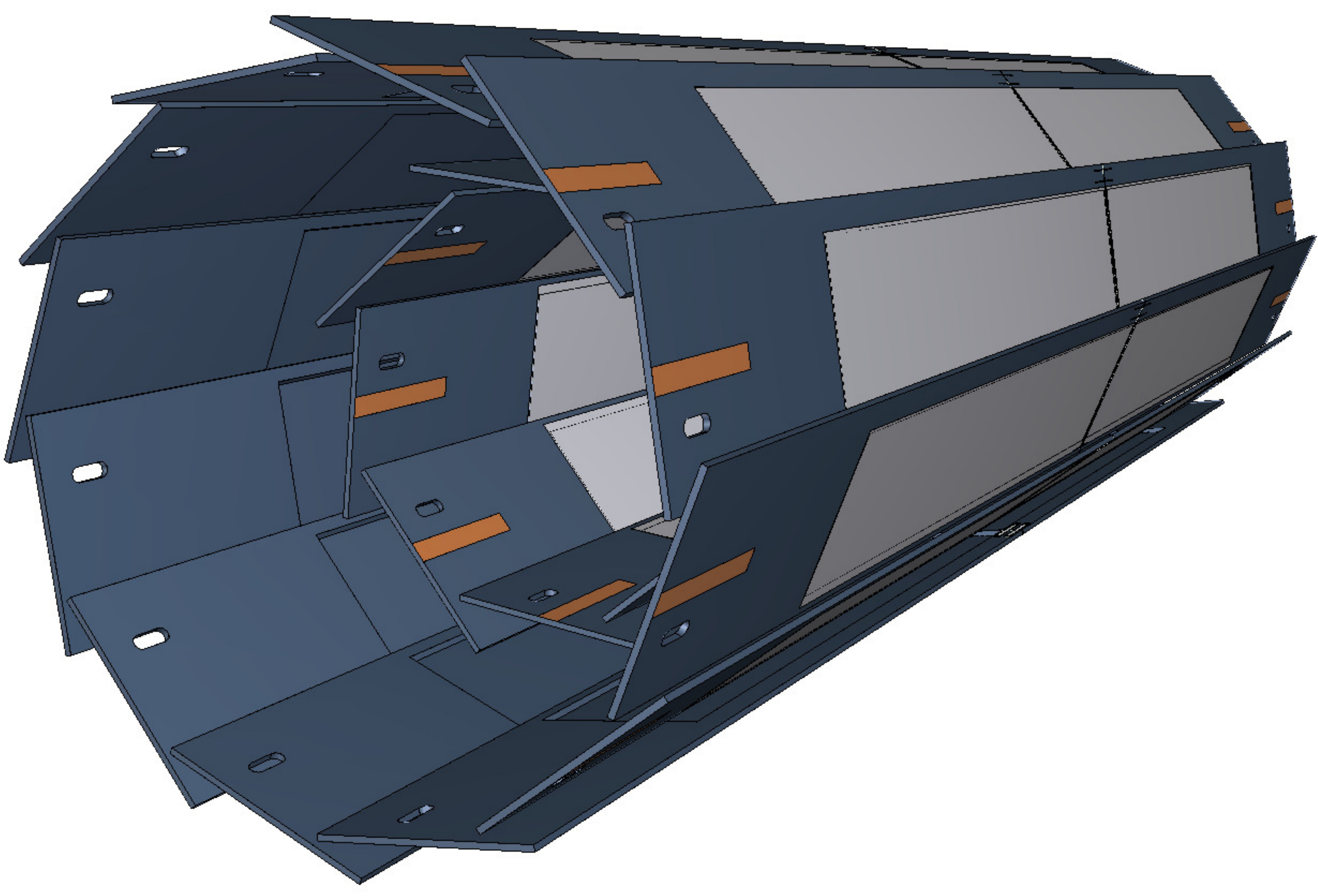}
\caption{A sketch of the Belle II pixel detector. Each ladder consists of
two half ladders and the inner layer has 8 layers while the outer one
has 12 layers. }
\label{fig:PXDDetector}
\end{figure}

The detector
consists of two cylindrical layers, with the innermost layer located 1.4 cm and the outer one 
2.2 cm away from the interaction point. The modules in the inner (outer) layer
contain small pixels of size $50 \times 55 \, \mu$m$^{2}$ ($50 \times 60 \, \mu$m$^{2}$)
and large pixels of size $50 \times 70 \, \mu$m$^{2}$ ($50 \times 85 \, \mu$m$^{2}$). Overall
the PXD is built out of about 8 million pixel cells, arranged in 40 half-ladders, $2\times8$ located
in the inner layer and $2\times12$ arranged in the outer one.  
Two half-ladders are front-face glued next to each other.
A sketch of the Belle II pixel
detector is shown in Fig.~\ref{fig:PXDDetector}.
The sensitive area of
the sensors will be thinned down to a thickness of 75 $\mu$m. For this the
backside of the sensor is processed and then bonded to a so-called
handling wafer. The sensor is then thinned down to the final thickness of 75 $\mu$m 
and the required processing steps are performed on the wafer. Finally the parts of the 
handling wafer are etched away and only a frame will remain.This
thinning process will result in 
an average material budget of the first layer of about $0.2\%$ X$_{0}$, dominated 
by the DEPFET sensor and the frame of the module, necessary for mechanical
stability. 
For this material budget we expect an average additional hit uncertainty 
of 5 $\mu m$ for a flight length of 1 cm (roughly the distance between
the first and second layer) and for a track of 1 GeV/$c$.\par
\begin{figure}[hbt] 
\centering 
\includegraphics[width=0.8750 \columnwidth,keepaspectratio]{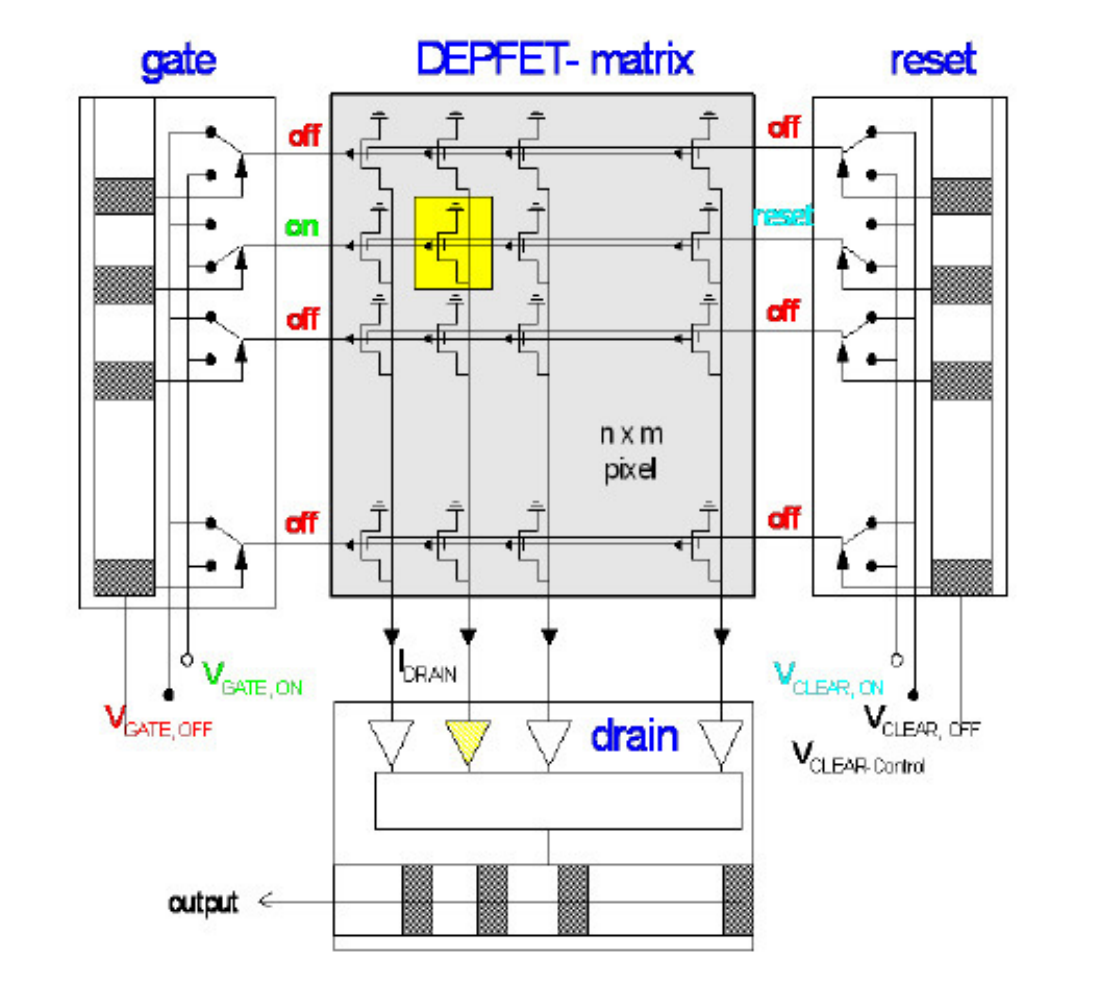}
\caption{Sketch illustrating the rolling shutter operation of the DEPFET matrix ~\cite{Abe:2010sj}.}
\label{fig:DEPFETMatrix}
\end{figure}

The PXD will be operated in the so-called rolling shutter mode. The pixels
are arranged in a grid and four pixel rows are read out simultaneously.
One half ladder contains 768 rows and the processing time for a single 
row is about 100 ns, dominated by the shaping time of about 70 ns.  
This leads to a read out time for a complete frame of 20 $\mu$s, which 
corresponds to two complete bunch cycles. This number has to be compared to
the expected event rate of about 30 KHz, which is completely dominated by 
background events. With the current design we expect an occupancy of $1 \%$ in the inner layer of the PXD. 
A sketch illustrating the readout scheme of the PXD is shown in Fig.~\ref{fig:DEPFETMatrix}.\par
\begin{figure}[hbt] 
\centering 
\includegraphics[width=0.875 \columnwidth,keepaspectratio]{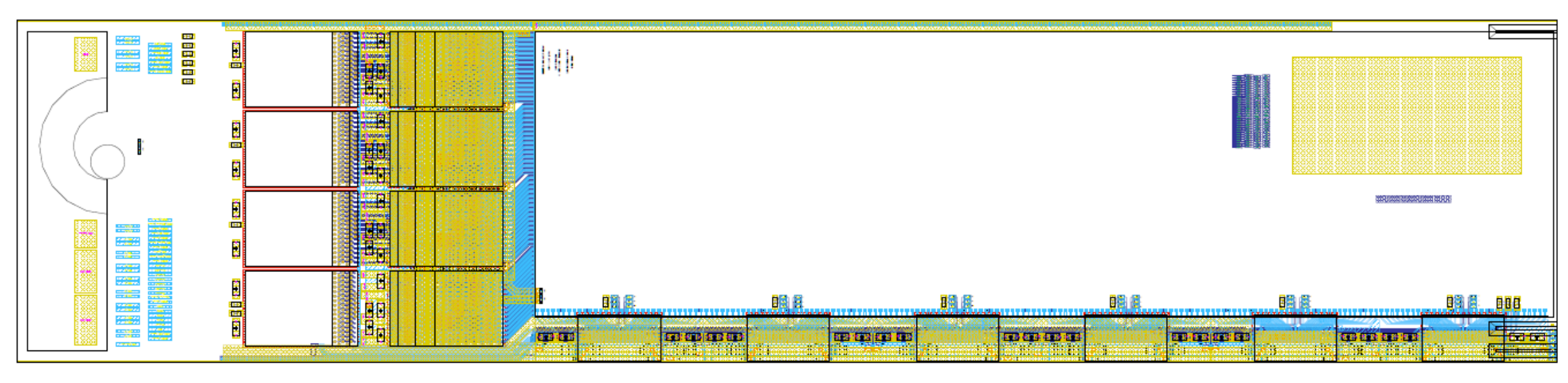}
\caption{Layout of a PXD half-ladder. The white large area corresponds
to the sensitive area, populated with $768\times256$ DEPFET pixels. 
The area below the sensitive area contains the six Switchers and 
left of the sensitive area the DCDB and the DHP can be found.}
\label{fig:DEPFETModule}
\end{figure}

For steering and reading out of the PXD three types of ASICs are operated 
on the all-silicon half-ladder. Each row of pixels is connected to a so-called 
Switcher, an ASIC responsible for selecting  pixels to be read out by applying a voltage to the 
external gate and to clear the internal gate after the readout has taken place. 
Since four rows are selected at the same time and a single Switcher can address 
32 channels at the same time six Switcher are needed for each half ladder.
The current from the drain of the selected DEPFET cells is 
processed in the DCDB, a dedicated ASIC responsible for
digitising the signal. In addition the DCDB also performs a common mode subtraction.
The digitised and common mode corrected signal is
then further processed by the DHP. This ASIC performs pedestal subtraction and zero suppression.
For each half ladder four DCDBs and DHPs
are needed for the 250 columns of DEPFET cells. The layout of a single PXD half-ladder is shown 
in Fig.~\ref{fig:DEPFETModule}.
The zero suppressed data coming from a single half module has a data rate of about 
4 Gbit/s and is transmitted via a 40 cm long kapton flex  outside the inner detector 
region of Belle II. Besides the transmission of data the kapton flex is also part of the
power distribution of the PXD. Each half module needs about 20 different voltages 
to supply the three different ASICs and the sensor. Outside the inner detector region
the kapton flex is replaced by an about 15 m long cables to connect the PXD
to the power supply modules~\cite{Rummel:2013nv} and to the data handling hybrid (DHH). The DHH is
responsible for serialising the data stream and to manage the trigger and
timing information. The data from all half-modules leads to a
data rate of 58 Gbit/s~\cite{Abe:2010sj}, a size which would by far dominate the
overall Belle II data rate. To further reduce the data rate originating from
the PXD a further data suppression scheme is applied. Tracks online reconstructed 
from hits in the SVD are extrapolated to the PXD and only hits in the PXD
associated to a certain region of interest defined by the SVD track are 
passed to the Belle II data acquisition. A significant fraction of hits originating
from background events, mainly from low energy QED processes, can be
rejected at an early stage.
\subsection{DEPFET Test Beam Measurements}
\begin{figure}[hbt] 
\centering 
\includegraphics[width=0.875 \columnwidth,keepaspectratio]{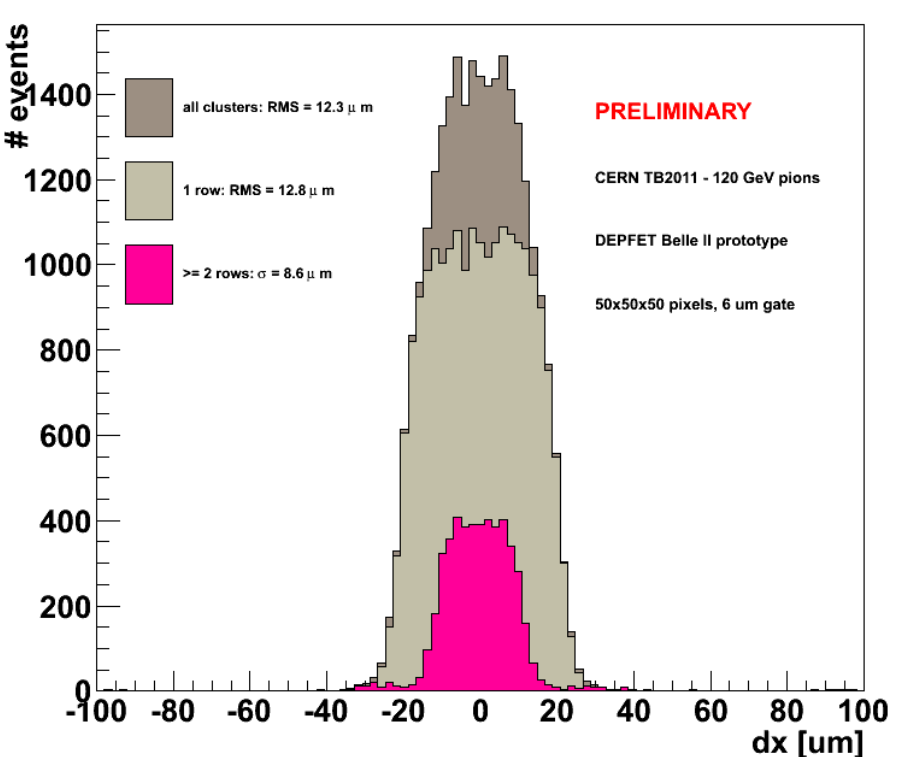}
\caption{Residual distribution of hits collected with a DEPFET matrix during a test beam. 
The red distribution corresponds to hits with a signal in more than two pixel rows, while
the light grey distribution corresponds to hits with a signal in a single row only. The
dark grey distribution is the sum of both~\cite{TestBeam}.}
\label{fig:SingleHitResolution}
\end{figure}

Dedicated tests of DEPFET devices are performed with test beam measurements
at CERN and DESY~\cite{TestBeam}. The design of the DEPFET matrices used in 
these tests is not identical to the final one, but close enough in order to
demonstrate the proof of principle.  
Full size matrices have been produced, but are not yet equipped with the ASICs. 
For the beam tests smaller matrices with 32 x 64 pixels, wire bonded to special PCBs carrying the ASICs have been used. 
The pixel size is different to the final design with  a size of $50\times50 \,\mu$m$^{2}$ and 
$50\times75 \,\mu$m$^{2}$. The sensor thickness was thinned down to 50 $\mu$m. 
Using the ASICs for steering the matrices and reading 
out the DEPFET pixels the design readout time of 100 ns per pixel row was achieved. The hit
efficiency achieved with this setup was more than $99\%$. A preliminary
result of the single hit resolution is shown in Fig.~\ref{fig:SingleHitResolution}
~\cite{TestBeam}. For the DEPFET pixels with a size of $50\times50 \,\mu$m$^{2}$
an intrinsic resolution of $12\, \mu$m (RMS) was measured. 
The signal-to-noise
ratio varies between 20 and 40, depending on the actual design of the DEPFET 
pixel cell.
\subsection{Radiation Hardness of the Detector}
\begin{figure}[hbt] 
\centering 
\includegraphics[width=0.875 \columnwidth,keepaspectratio]{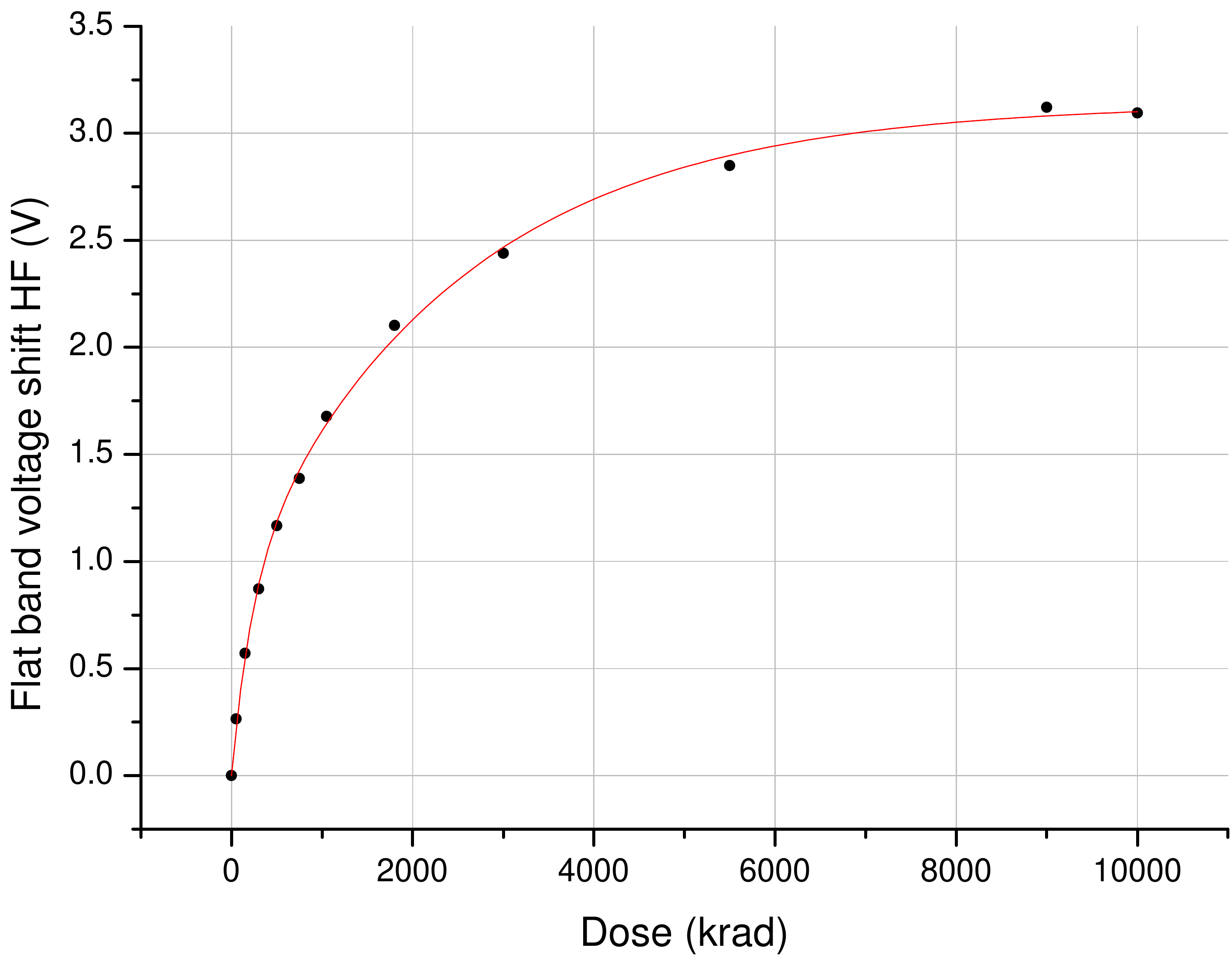}
\caption{Flat band voltage shift of a MOSFET as a function of radiation dose. The 
MOSFET contains a thin SiO$_{2}$ layer (75 nm) and an additional Si$_{3}$N$_{4}$ layer~\cite{RadHard}.}
\label{fig:MOSFETRadiation}
\end{figure}

The most dangerous background events during the operation of the 
Super KEKB-accelerator are low momentum electron-positron pairs originating from
QED processes. Two different irradiation damage mechanisms can affect
the operation of the PXD. Irradiation can cause damage of the silicon 
bulk material as well as creation of additional surface charges or oxide damage 
being located close to material boundaries. The two different damage types
lead to different problems during the operation of the PXD.
Bulk damage induces an increased leakage current with the
produced charged particles ending up in the internal gate 
of the DEPFET. Charges created by the primary ionising particle 
can be trapped or de-trapped, leading to changes in the signal current. Finally, 
in case of  a very large flux, the effective doping of the device could change. 
However, the expected flux of $\Phi_{\mathrm{eq}}= 1.2 \times 10^{13}$ cm$^{-2}$
for ten years of operation, assuming a hardness of electrons of 0.06, which is based 
on the NIEL scaling assumption, is not expected to compromise
the operation of the PXD. 
On the contrary surface charges and oxide damages caused by 
irradiation can have
a significant impact on the PXD operation. Positive charges collected close
to the Si/SiO$_{2}$ boundary lead to a shift of the gate and clear operation voltage 
towards more negative values.
First tests using DEPFET matrices developed
for the  future linear accelerator indicated a shift in the threshold voltage
of more than 10~V during the lifetime of the PXD. A nonuniform
irradiation of modules could lead to significantly different operation 
voltages within a single half-module. With the current design, 
running six switchers on a single half-module, only a limited
number of different operation voltages can be applied to the individual pixels
and a nonuniform irradiation can not be 
counterbalanced by changing the operation voltage. To reduce the
sensitivity to these types of irradiation effects the outline of the DEPFET pixel
cell is modified. Besides the reduction of the SiO$_{2}$ layer of the DEPFET cell, 
leading to a reduced voltage shift, an additional Si$_{3}$N$_{4}$ layer
is incorporated which accumulates negative charges which in addition partly 
compensate the negative voltage shift. The flat band voltage change of a MOSFET structure
similar to the one used for the DEPFET as a function of radiation dose is shown in 
Fig.~\ref{fig:MOSFETRadiation}~\cite{RadHard}. The expected dose 
for one year of operation is 1.85 MRad/yr and Fig.~\ref{fig:MOSFETRadiation} 
indicates that the voltage shift due to irradiation is significantly reduced
with no impact on the operation of the PXD.

\subsection{Thermal Management}

\begin{figure}[hbt] 
\centering 
\includegraphics[width=0.875 \columnwidth,keepaspectratio]{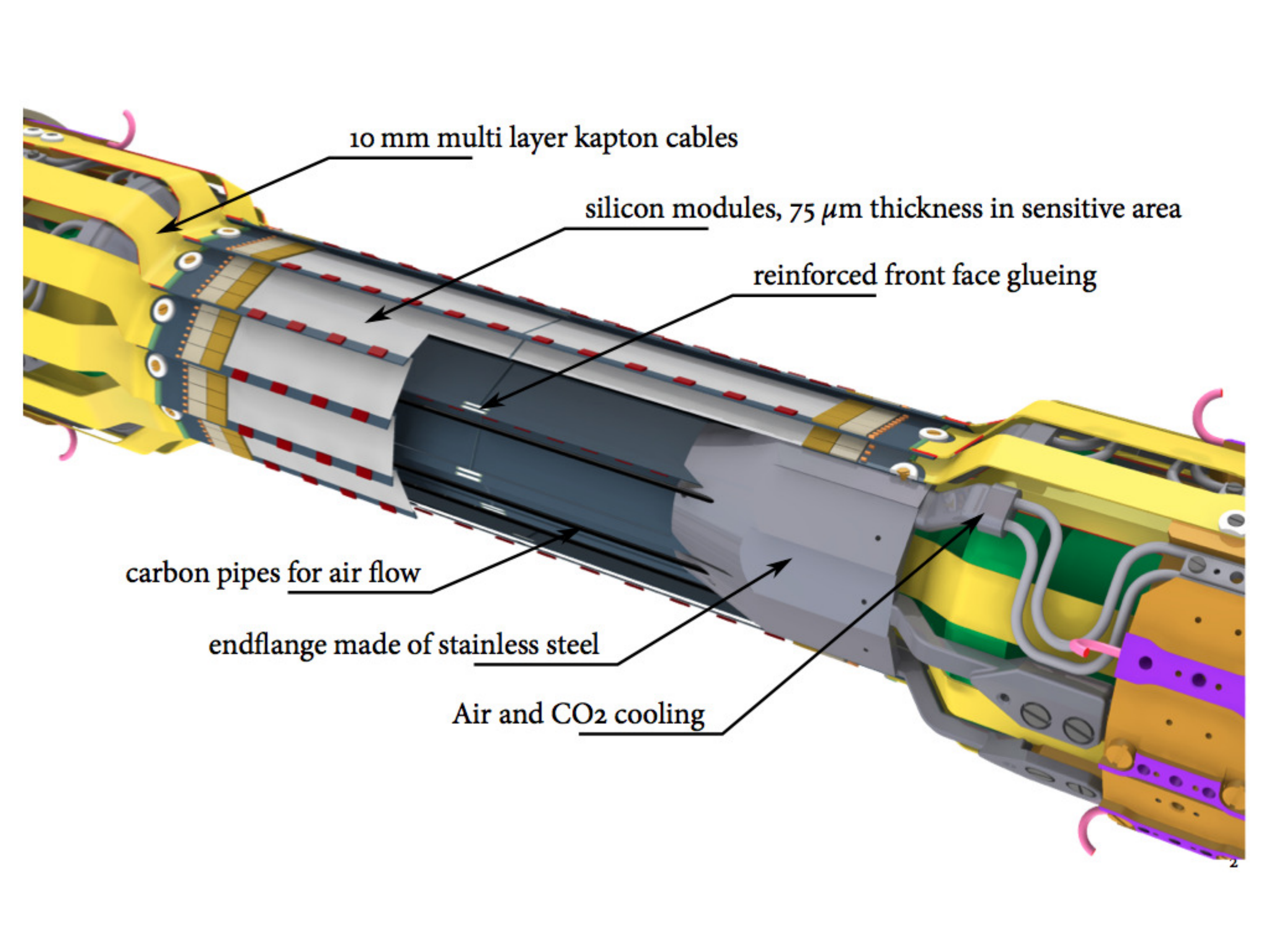}
\caption{Overview of the cooling concept used for the PXD. The ASICs located outside
the acceptance region are mounted on CO$_{2}$ cooled end-flange and the switchers and 
the sensor is cooled by cold air.}
\label{fig:DEPFETCooling}
\end{figure}

The low power consumption of the PXD is one of the major advantages of the DEPFET design. 
As already mention in section~\ref{depfetprinciple} the power consumption from the active sensor
is about 0.5 W per half ladder. The major contribution to the total power consumption is produced by
the ASICs located next to the sensitive area. During operation a power dissipation of
360 W is expected for the entire PXD, dominated by the readout ASICs. 
The DCDB and DHP are placed outside the acceptance where the ladder with the ASICs is mounted with
direct thermal contact on 
a CO$_{2}$ cooled mounting block. The switchers, being distributed along the ladder,
are cooled by an air flow, with a cooling power of about 100 mW/cm$^{2}$.
See Fig.~\ref{fig:DEPFETCooling} for an overview of the
PXD cooling concept.

\section{Expected Performance of the Belle II Vertex Detector}

\begin{figure}[hbt] 
\centering 
\includegraphics[width=0.875 \columnwidth,keepaspectratio]{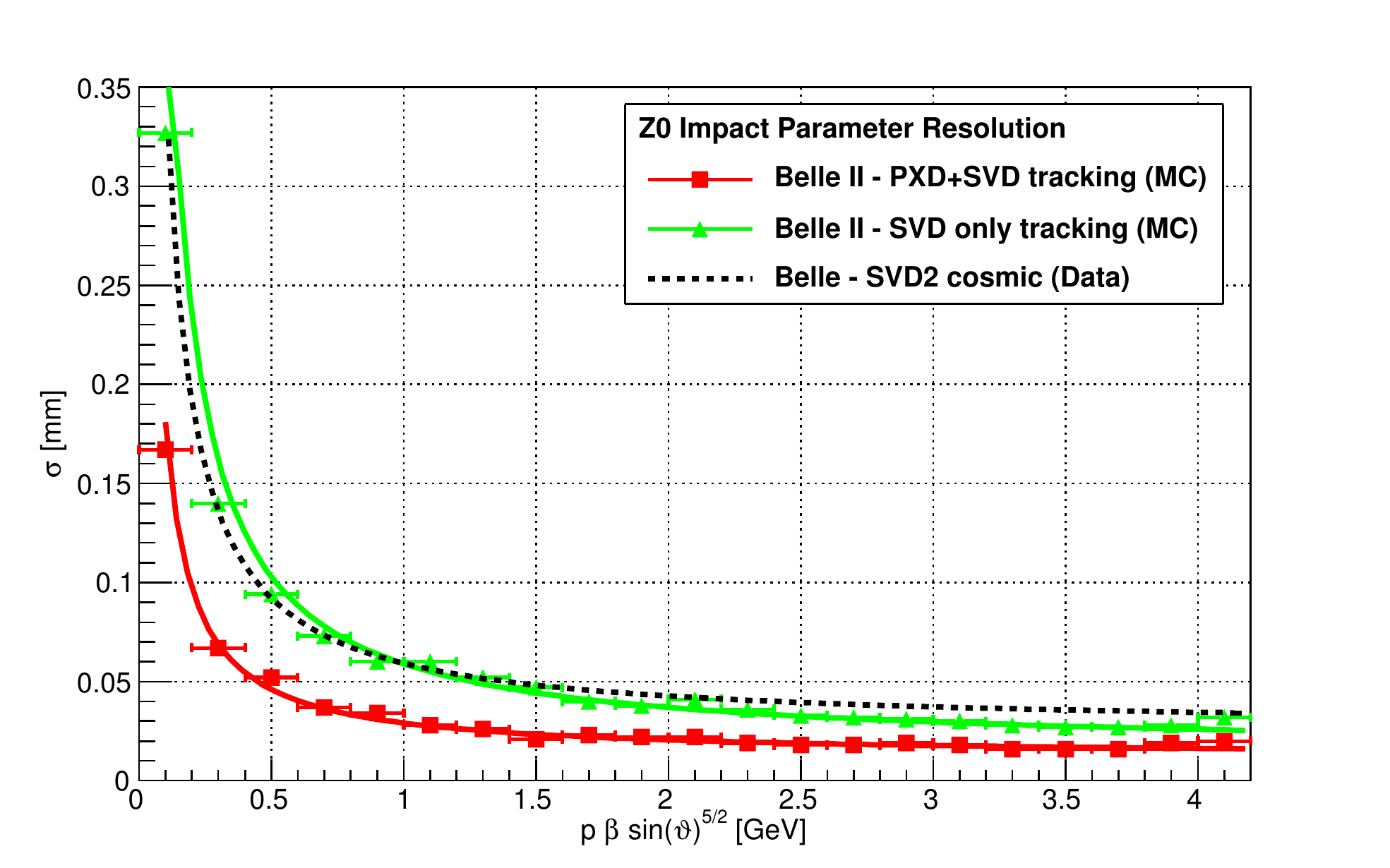}
\caption{Resolution of the $z_{0}$-parameter of charged particles as 
a function of the momentum scaled with respect to the incident angle, 
estimated with simulated events. The green points correspond to 
a simulation with SVD tracking only, the red points to a simulation 
with SVD and PXD tracking and the dashed line is from a measurement
using cosmic tracks collected with the Belle experiment ~\cite{Drasal:2012zh}.  }
\label{fig:z0resolution}
\end{figure}

The expected performance is estimated using simulated events~\cite{Drasal:2012zh}. Charged
particles reconstructed with Belle II can be described with helices and 
the PXD will mainly contribute to the impact parameter determination, $d_{0}$ and $z_{0}$, 
with respect of the interaction point. Fig.~\ref{fig:z0resolution} shows the
resolution of the $z_{0}$-parameter, the distance between the charged particle
and the interaction point in the direction along the incoming beams, for different detector layouts. 
Reconstruction of charged particles  in Belle II, with a vertex detector
only consisting of a silicon strip detector, would lead to a similar resolution
as for predecessor experiment Belle. Using in addition the information from the PXD
will, as expected, significantly improve the tracking 
performance of the experiment. The intrinsic hit resolution of the PXD will 
lead to a gain in resolution for high momentum tracks, while the very low
material budget will significantly improve the performance for low
momentum charged particles. The study returns a similar conclusion 
for the resolution of the $d_{0}$-parameter.

\section{Summary and Conclusion}

Measurements using the Belle II-detector at the Super KEKB-accelerator offer
an orthogonal approach for searches for physics beyond the SM. To
exploit the full physics potential the detector is equipped with a low-material
pixel detector close to the interaction point. The detector is based on the
DEPFET-principle which fulfils all requirements for a successful data taking and
physics analysis at the high luminosities delivered by Super KEKB
The PXD is expected to be ready for first data taking in 2015.




\end{document}